\documentclass[aip,jcp,reprint,amsmath,amssymb]{revtex4-2}

\usepackage[version=4]{mhchem}       
\usepackage{graphicx}                
\usepackage{booktabs}                
\usepackage{threeparttable}          
\usepackage{microtype}               

\usepackage{dcolumn}
\usepackage{bm}

\usepackage[utf8]{inputenc}
\usepackage[T1]{fontenc}
\usepackage{mathptmx}
\usepackage{etoolbox}
\DeclareUnicodeCharacter{223C}{\ensuremath{\sim}}
\DeclareUnicodeCharacter{03B1}{\ensuremath{\alpha}}
\DeclareUnicodeCharacter{03B3}{\ensuremath{\gamma}}
\makeatletter
\let\JA@orig@bibinfo\bibinfo
\def\JA@journal@name{journal}
\newcommand{\DefJournalAbbrev}[2]{%
  \expandafter\gdef\csname JA@@\detokenize{#1}\endcsname{#2}}
\renewcommand{\bibinfo}[2]{%
  \def\JA@this{#1}%
  \ifx\JA@this\JA@journal@name
    \@ifundefined{JA@@\detokenize{#2}}{#2}{\csname JA@@\detokenize{#2}\endcsname}%
  \else
    \JA@orig@bibinfo{#1}{#2}%
  \fi}
\makeatother

\DefJournalAbbrev{The Journal of Chemical Physics}{J. Chem. Phys.}
\DefJournalAbbrev{The Journal of Physical Chemistry A}{J. Phys. Chem. A}
\DefJournalAbbrev{The Journal of Physical Chemistry B}{J. Phys. Chem. B}
\DefJournalAbbrev{The Journal of Physical Chemistry C}{J. Phys. Chem. C}
\DefJournalAbbrev{The Journal of Physical Chemistry Letters}{J. Phys. Chem. Lett.}
\DefJournalAbbrev{Physical Review Letters}{Phys. Rev. Lett.}
\DefJournalAbbrev{Physical Review B}{Phys. Rev. B}
\DefJournalAbbrev{Physical Chemistry Chemical Physics}{Phys. Chem. Chem. Phys.}
\DefJournalAbbrev{Journal of Radioanalytical and Nuclear Chemistry Letters}{J. Radioanal. Nucl. Chem. Lett.}
\DefJournalAbbrev{Journal of Nuclear Materials}{J. Nucl. Mater.}
\DefJournalAbbrev{Journal of Hazardous Materials}{J. Hazard. Mater.}
\DefJournalAbbrev{Journal of Computational Chemistry}{J. Comput. Chem.}
\DefJournalAbbrev{International Journal of Hydrogen Energy}{Int. J. Hydrogen Energy}
\DefJournalAbbrev{Geophysical Research Letters}{Geophys. Res. Lett.}
\DefJournalAbbrev{Environmental Science \& Technology}{Environ. Sci. Technol.}
\DefJournalAbbrev{Computational Materials Science}{Comput. Mater. Sci.}
\DefJournalAbbrev{Chemical Physics Letters}{Chem. Phys. Lett.}
\DefJournalAbbrev{Applied Surface Science}{Appl. Surf. Sci.}
\DefJournalAbbrev{Applied Spectroscopy}{Appl. Spectrosc.}
\DefJournalAbbrev{ACS Applied Nano Materials}{ACS Appl. Nano Mater.}
\DefJournalAbbrev{Thermochimica Acta}{Thermochim. Acta}
\DefJournalAbbrev{Surface Science}{Surf. Sci.}
\def\selectlanguage#1{}
\makeatother
\usepackage{siunitx}

\newcommand{\orcid}[1]{}

\begin{document}

\title{%
  Electron-Stimulated Desorption of D Atoms from Gibbsite
  (\ce{Al(OD)3}) and \ce{D2O} Ice: Energy and Temperature
  Dependence of Translational Energy Distributions
}

\author{William T. P. Denman}
  \orcid{0000-0003-4752-0073}
  \affiliation{School of Chemistry and Biochemistry,
    Georgia Institute of Technology, Atlanta, Georgia 30332, USA}

\author{Brant M. Jones}
  \orcid{0000-0002-6704-1064}
  \affiliation{School of Chemistry and Biochemistry,
    Georgia Institute of Technology, Atlanta, Georgia 30332, USA}

\author{Jacob Messner}
  \orcid{0000-0002-8199-6019}
  \affiliation{School of Chemistry and Biochemistry,
    Georgia Institute of Technology, Atlanta, Georgia 30332, USA}

\author{Xin Zhang}
  \orcid{0000-0003-2000-858X}
  \affiliation{Physical Sciences Division,
    Pacific Northwest National Laboratory, Richland, Washington 99352, USA}

\author{Micah P. Prange}
  \orcid{0000-0002-0775-2024}
  \affiliation{Physical Sciences Division,
    Pacific Northwest National Laboratory, Richland, Washington 99352, USA}

\author{Greg A. Kimmel}
  \orcid{0000-0001-9766-5399}
  \affiliation{Physical Sciences Division,
    Pacific Northwest National Laboratory, Richland, Washington 99352, USA}

\author{Jay A. LaVerne}
  \orcid{0000-0002-8383-7476}
  \affiliation{Radiation Laboratory and Department of Physics
    \& Astronomy, University of Notre Dame,
    Notre Dame, Indiana 46556, USA}

\author{Thomas M. Orlando}
  \orcid{0000-0002-2422-4506}
  \email{thomas.orlando@chemistry.gatech.edu}
  \affiliation{School of Chemistry and Biochemistry,
    Georgia Institute of Technology, Atlanta, Georgia 30332, USA}
    \affiliation{School of Physics, Georgia Institute of Technology, Atlanta, Georgia 30332, United States;}

\date{\today}

\begin{abstract}
The electron-stimulated desorption (ESD) of neutral D atoms from gibbsite (\ce{Al(OD)3}) nanoplatelets and amorphous \ce{D2O} ice has been investigated using $2+1$ resonance-enhanced multiphoton ionization (REMPI) time-of-flight mass spectroscopy in a high vacuum chamber at temperatures 15 and 300\,K. Electron irradiation at 540, 250, and 150\,eV 
produces similar translational energy distributions at $\sim$300\,K, with a dominant intermediate-temperature component ($T \sim 1500$--$2100$\,K). Cooling to 15\,K suppresses the D atom yield by approximately 50\% and removes the lowest-temperature (slowest) component. This decrease in yield is consistent with diminished hole mobility and restricted diffusion at cryogenic temperatures. Under identical conditions, \ce{D2O} amorphous solid water ice films produce approximately 20 times greater D atom signal than bare gibbsite, with significantly hotter translational distributions, reflecting the higher deuterium surface density and distinct bonding environments of bulk ice relative to the terminal hydroxyl groups on gibbsite. These results identify hole transport to terminal hydroxyl sites as the rate-limiting step for nonthermal D atom production and provide a mechanistic framework for understanding atomic hydrogen release from aluminum hydroxide phases relevant to radioactive waste storage at the Hanford Site.

\end{abstract}

\maketitle

\section{Introduction}

Gibbsite (\ce{Al(OH)3}) plays a critical role as an ore of aluminum and is one of the key components of the aggregated solid waste present in the radioactive tank waste stored at the Department of Energy’s legacy nuclear site, Hanford.\cite{petersonReviewScientificUnderstanding2018} Since gibbsite can persist in these highly radioactive environments, there have been numerous studies examining the role these particles may have in the nonthermal production of molecular hydrogen, \ce{H2}, and oxygen, \ce{O2}, upon exposure to  ionizing radiation .\cite{jonesElectronStimulatedFormationRelease2022,huestisRadiolyticStabilityGibbsite2018,kaddissy_radiolytic_2017} Consequently, the production of these gases which can be trapped in waste sludge and then released is of significant importance due to the related safety concerns with their production. Studies have shown that the dominant product during radiolysis of boehmite and gibbsite is \ce{H2}. A stepwise mechanism was proposed by Kaddissy et al. to explain the production of \ce{H2}. \cite{kaddissy_efficient_2019} They suggested that \ce{H^{.}} atoms are formed from dissociation of the hydroxyl group. The \ce{H^{.}} atoms can then diffuse and react to form \ce{H2}, i.e., \ce{H^{.} + H^{.} -> H2}, or undergo abstraction reactions \ce{H^{.} + OH^- -> H2(g) + O^{.-}}. \cite{jonesElectronStimulatedFormationRelease2022} In addition, some of these \ce{H^{.}} atoms can become trapped and require an activation energy for removal as has been verified recently. \cite{wangRoleSurfaceHydroxyls2020} EPR data has shown the creation of \ce{H2} in boehmite and gibbsite with irradiation and subsequent heating.\cite{kaddissy_radiolytic_2017} Kaddissy et al \cite{kaddissy_radiolytic_2017} attribute the location of one of the observed radiation induced defects to a trapped hole in the filled 2p orbital of oxygen. A later computational study suggested that the \ce{O-} diffusion occurred via proton-coupled hole transfer.\cite{shenDiffusionMechanismsRadiolytic2018} 
Thus, the difference in radiolytic yields are likely associated with hole transport to the surface. This is consistent with previous observations that hole mobility to the surface was shown to limit the rate of water oxidation in the chemically similar mineral (hematite, \ce{\alpha-Fe2O3}).\cite{prange_electronic_2018} In this report, the formation of \ce{D} neutral species from gibbsite during low energy electron scattering is observed with REMPI and time-of-flight spectrometry. This is compared with \ce{D} production from \ce{D2O} ice at 15\,K and further computational studies.

\section{Experimental and Data Analysis}

\subsection{Laboratory Measurements}

Irradiation experiments were performed in a custom-built high vacuum (HV) chamber with a base pressure of $\sim\!1\times10^{-9}$\,Torr equipped with a homebuilt time-of-flight (TOF) mass spectrometer, as described previously. \cite{jonesElectronThermalStimulatedSynthesis2019,siegerElectronstimulatedDesorption21997,desimoneMechanismsH2ODesorption2013} Samples were mounted on a closed-cycle helium cryostat (ArcCryo) capable of maintaining temperatures between 15 and 330\,K. Electron irradiation was performed using a tunable-energy (5--1000\,eV) electron gun (Kimball Physics, ELG-2) operating with a 500\,ns pulse width. The rising edge of the extractor voltage pulse coincided with the end of the electron gun pulse, defining the time origin for the TOF measurement.
Three incident electron energies were employed: 540, 250, and 150\,eV. These energies were chosen to probe distinct electronic excitation regimes of the gibbsite target surface. The O\,1$s$ binding energy in gibbsite is approximately 532\,eV~\cite{zhang_boehmite_2018}, so 540\,eV irradiation accesses O\,1$s$ core-hole excitations and the resulting Auger cascade, while 250 \,eV and 150 \,eV are below this threshold allowing direct comparison of the O\,1$s$ contribution. Both 250 and 150\,eV exceed the Al\,2$p$ and Al\,2$s$ binding energies ($\sim$74 and $\sim$118\,eV, respectively).~\cite{zhang_boehmite_2018} All three energies are well above the valence excitation thresholds for electron-stimulated desorption of \ce{D+} from hydroxyl-bearing surfaces~\cite{siegerElectronstimulatedDesorption21997}. 
\par
Desorbing neutral D atoms were detected by $2+1$ resonance-enhanced multiphoton ionization (REMPI). The probe laser system consisted of a Nd:YAG laser (532\,nm) pumping a Continuum ND-6000 dye laser, producing a fundamental output at 615.264\,nm. This beam was directed through an InRad Autotracker~III harmonic generation unit, producing the third harmonic at 205.09\,nm via nonlinear wave mixing. This wavelength corresponds to the two-photon resonance of ground-state D atoms at the $1s \to n\!=\!3$ transition. Ions produced in the laser focal volume were accelerated into the TOF mass spectrometer and detected by a microchannel plate (MCP) detector (5\,$\times$\,435 FT-400).
The delay time between the end of the electron pulse and the ionization laser pulse was stepped incrementally using a delay generator, mapping the arrival-time distribution of desorbing D neutral species to the ionization region. Simultaneously, the electron-stimulated desorption (ESD) ion signal was recorded at each delay step as a diagnostic of electron beam alignment and sample activity.~\cite{siegerElectronstimulatedDesorption21997}

\subsection{Sample Preparation}

Gibbsite (\ce{Al(OH)3}) nanoplatelets were synthesized at Pacific Northwest National Laboratory (PNNL) following an additive-free hydrothermal protocol.~\cite{zhang_boehmite_2018} The as-synthesized material consists of hexagonal nanoplates with average lateral dimensions of $\sim$280\,nm and thickness of $\sim$18\,nm, with dominant (001) basal surfaces terminated by hydroxyl groups.~\cite{zhang_boehmite_2018}

Samples were prepared by dispersing 0.1\,g of gibbsite nanoplatelets in 1\,mL of \ce{D2O} (Sigma-Aldrich) to form a uniform slurry. The suspension was agitated and then deposited via pipette onto a polished oxygen-free copper sample disc. The slurry was dried on a hot plate at $\sim$330\,K until all liquid had evaporated. The sample disc was then mounted vertically to the cryostat cold head. The sample was held at 300\,K during pump down and then held at
($\sim\!1\times10^{-9}$\,Torr) to remove residual surface water before irradiation.

For substrate comparison experiments, thin films of \ce{D2O} ice were deposited onto the gibbsite surface \textit{in situ} via Langmuir dosing through a leak valve with the sample held at 15\,K. Film coverage in monolayers (ML) was estimated from the backing pressure of the \ce{D2O} source, the sample temperature, and the deposition time using the Langmuir exposure relation (1\,L $= 1\times10^{-6}$\,Torr\,s, assuming unity sticking probability at 15\,K). The angle of deposition between the inlet and the surface was between 0\textdegree and 40\textdegree thus producing non-porous ASW. \cite{stevensonControllingMorphologyAmorphous1999}

Time-of-flight (TOF) mass spectrometry data were collected with raw transient signals recorded with varied delay step as digitized voltage traces. For each scan, the \ce{D} atom REMPI signal was extracted by numerically integrating over a predefined time window using the trapezoidal rule. 
ESD ion signal was extracted analogously from a separate time window corresponding to hydrogen ion arrival. This peak was used for both timing and calibration but an ion velocity distribution was not extracted due to experimental constraints.

Two complementary signal metrics were computed for each delay step: the peak amplitude (maximum voltage within the integration window) and the integrated area. The integrated area was used as the primary signal quantity for all subsequent fitting. A summary of signal characteristics across all experimental conditions is provided in Table~\ref{tab:signal_summary}.

\begin{table*}[htbp]
\centering
\caption{Summary of D atom REMPI signal characteristics for all experimental conditions. Peak arrival time and integrated peak area are reported as mean $\pm$ standard deviation across scans within each condition group.}
\label{tab:signal_summary}
\begin{tabular}{lccc}
\toprule
Condition
  & Peak arrival ($\mu$s)
  & D atom area (arb.)
  & ESD area (arb.) \\
\midrule
\multicolumn{4}{l}{\textit{Gibbsite, 300\,K}} \\[2pt]
\quad 540\,eV 300\,K & $0.300 \pm 0.079$ & $6.709 \pm 0.482$ & $33.229 \pm 8.875$  \\
\quad 250\,eV 300\,K & $0.162 \pm 0.060$ & $4.299 \pm 1.455$ & $19.017 \pm 10.197$  \\
\quad 150\,eV 300\,K & $0.137 \pm 0.032$ & $5.880 \pm 0.314$ & $61.401 \pm 0.449$  \\
\\[4pt]
\multicolumn{4}{l}{\textit{Gibbsite, cryogenic}} \\[2pt]
\quad 250\,eV 15\,K & $0.110 \pm 0.042$ & $2.172 \pm 0.212$ & $32.821 \pm 7.455$  \\
\\[4pt]
\multicolumn{4}{l}{\textit{\ce{D2O} ice, 15\,K}} \\[2pt]
\quad 250\,eV 15\,K & $0.120 \pm 0.028$ & $42.407 \pm 7.917$ & $75.185 \pm 2.193$  \\
\\[4pt]

\bottomrule
\end{tabular}
\end{table*}



\subsection{Maxwell-Boltzmann Velocity Distribution}

Though nonthermal desporption is not an equilibrium process, the resulting velocity distributions are often fit to a Maxwell-Boltzmann distribution with the effective temperature being analogous to the energy gained from the repulsive upper state potential. \cite{zimmermannVelocityDistributionsPhotochemically1994} The sample-to-ionization-volume distance $r$, defined as the perpendicular distance between the sample surface and the laser focal point, enters directly into the Maxwell-Boltzmann velocity distribution model through the relation $v = r / (t - t_0)$, where $t$ is the measured delay time and $t_0$ is the neutral formation time. Accurate determination of $r$ is therefore critical for obtaining physically meaningful fitted temperatures. The calibrated distance for each experimental run was calculated as:

\begin{equation}
    r = r_\mathrm{ref} \times \frac{t}{t_\mathrm{ref}}
    \label{eq:rdist_cal}
\end{equation}

The measured D atom delay-time distributions were modeled as a sum of Maxwell-Boltzmann (MB) number density distributions. Under the number density condition, in which the laser fired at fixed delay time ionizes the atoms in the focal volume at that instant, the signal at delay time $t$ is given by:

\begin{equation}
    S(t) = \sum_{i} A_i \, (t - t_0)^{-3} \exp\!\left[ \frac{-m r^2}{2 k_\mathrm{B} T_i (t - t_0)^2} \right]
    \label{eq:mb_flux}
\end{equation}

where $m$ is the mass of a D atom, $k_\mathrm{B}$ is the Boltzmann constant, $r$ is the calibrated flight distance, $t_0$ is the neutral formation time, and $A_i$ and $T_i$ are the amplitude and characteristic temperature of the $i$-th component. Fitting was performed in the time domain using nonlinear least-squares optimization. Both two-component (2MB) and three-component (3MB) models were evaluated for each experimental condition.

The neutral formation time, $t_0$, was constrained to the interval $[-300, -200]$\,ns, corresponding center of the gaussian distribution for the duration of the 500\,ns electron gun pulse. This bounded optimization avoids nonphysical solutions in which $t_0$ falls outside the pulse window while retaining sensitivity to its true value within the physically reasonable range. Due to geometry and experimental constrains, the electron beam and the extractor field can not be operated congruently. The sensitivity of the fitted temperatures to the choice of $t_0$ within this interval was assessed and is reported in the Supplementary Information.

The value $r = 3.60$\,mm was physically measured using a calibrated translation stage and serves as the primary reference. Because the laser focal point position was adjusted between experimental runs, $r$ cannot be assumed constant across the full data set. A calibration procedure was therefore applied to determine $r$ for each experimental run independently.

Model selection between the 2MB and 3MB fits was performed using the Akaike Information Criterion (AIC) and Bayesian Information Criterion (BIC). Both criteria penalise model complexity to guard against overfitting: the 2MB model has five free parameters (two amplitudes, two temperatures, and $t_0$) while the 3MB model has seven. The preferred model was taken to be that with the lower BIC value. In cases where the BIC difference was less than two units, the simpler 2MB model was retained. An additional physical constraint was applied: a $T_3$ value converging to the upper optimisation bound of $10^6$\,K was taken as evidence that the three-component model was not justified, and the 2MB result was used instead.

The fractional contribution of each MB component to the total desorbing flux was calculated by numerically integrating each component over a fixed time grid spanning 50\,ns to 6\,$\mu$s using the trapezoidal rule, and dividing by the total integrated signal. These fractions represent the relative population of D atoms desorbing with each characteristic temperature and are reported alongside the fitted temperatures in the results.

\subsection{Density Functional Theory Calculations}
Plane wave density functional theory simulations of the electronic structure were performed within the Perdew-Burke-Ernzerhof (PBE) exchange-correlation approximation \cite{perdewGeneralizedGradientApproximation1996} using the PAW method \cite{blochlProjectorAugmentedwaveMethod1994}. The VASP code and associated PAW datasets, which left 3, 6, and 1 explicit valence electrons for Al, O, and H, respectively were used.\cite{kresseEfficiencyAbinitioTotal1996,PhysRevB.54.11169} The basis set was determined by an energy cutoff of 500\,eV. The DFT-D3 dispersion corrections of Grimme were employed. \cite{grimmeSemiempiricalGGAtypeDensity2006} Slab models exposing the basal (001) surface of gibbsite with 4 layers each containing 8 \ce{Al(OH)3} formula units were constructed. Periodic slab images were separated by 20 \AA{} of vacuum. The lateral dimensions of the orthorhombic slab models were fixed at 8.66 \AA{} $\times$ 10.08 \AA{} to correspond to the relaxed bulk. A $3 \times 3 \times 1$ k-point grid was used to sample the Brillouin zone, and occupations were smeared by 0.01 eV. A periodic bulk supercell, also containing 32 \ce{Al(OH)3} formula units ($a=13.2$, $b=10.1$, $c=12.4$ \AA{}; $\alpha=\gamma=90^{\circ}$, $\beta=95.29^{\circ}$) was considered as a reference. A $2 \times 3 \times 2$ k-grid was used for this model. Polarons (excitons) were simulated by restricting the spin state to contain two more spin-up electrons than spin-down electrons (i.e. a triplet). Atomic coordinates were optimized until all forces were less than 0.01\,eV/\AA{} in all models.
\section{Results}


\begin{table*}[htbp]
\centering
\caption{Maxwell-Boltzmann fit parameters for all experimental
conditions. Model selection (2MB or 3MB) was performed using
the Bayesian Information Criterion. The neutral birth time
$t_0$ was constrained to $[-300, -200]$\,ns}
\label{tab:mb_results}
\begin{tabular}{llcccccccc}
\toprule
Condition & Model
  & $T_1$ (K) & $f_1$ (\%)
  & $T_2$ (K) & $f_2$ (\%)
  & $T_3$ (K) & $f_3$ (\%)
  & $t_0$ (ns) \\
\midrule
\multicolumn{9}{l}{\textit{Gibbsite, 300\,K}} \\[2pt]
\quad 540\,eV 300\,K & 3MB & 186 & 20 & 2257 & 50 & 9343 & 30 & $-282$  \\
\quad 250\,eV 300\,K & 3MB & 186 & 19 & 1436 & 44 & 6331 & 37 & $-300$   \\
\quad 150\,eV 300\,K & 3MB & 149 & 20 & 1756 & 44 & 7963 & 35 & $-300$   \\
\\[4pt]
\multicolumn{9}{l}{\textit{Gibbsite, cryogenic}} \\[2pt]
\quad 250\,eV 15\,K & 2MB & 968 & 41 & 7231 & 59 & \multicolumn{2}{c}{---} & $-300$   \\
\\[4pt]
\multicolumn{9}{l}{\textit{\ce{D2O} ice, 15\,K}} \\[2pt]
\quad 250\,eV 15\,K & 2MB & 1985 & 43 & 6682 & 57 & \multicolumn{2}{c}{---} & $-278$  \\

\\
\end{tabular}
\end{table*}

\subsection{Electron Irradiation of Gibbsite: Energy Dependence at 300\,K}
Electron beam irradiation of bare gibbsite films at 300\,K showed distinct ESD peaks [Shown in SI figure] in our TOF  measurements as were two distinct laser induced peaks. The first peak ESD was attributed to protons and the first laser signal was correlated with REMPI of ground-state D atoms. The secondary ESD peak did not have a defined structure and due to the method of collection was unidentified. The second laser induced peak was very weak and has been attributed to non-resonant multiphoton ionization of neutral molecular oxygen.

\begin{figure}[htbp]
  \includegraphics[width=\columnwidth]{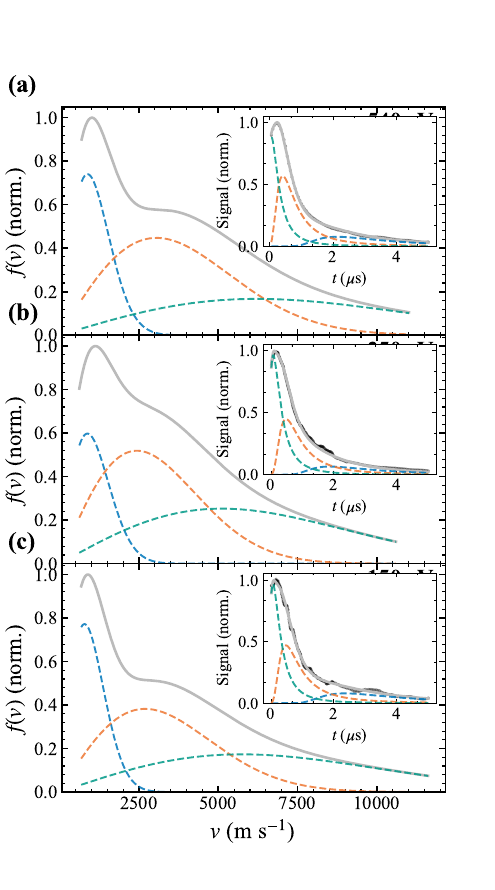}
  \caption{
    Velocity distributions of D atoms desorbed from Gibbsite at 300\,K for three electron irradiation energies. Each panel shows the normalized Maxwell-Boltzmann velocity three-component fit (solid grey line) and individual components $T_1$ (blue dashed), $T_2$ (orange dashed), and $T_3$ (teal dashed). Figure \textbf{a} represents irradiation at 540\,eV, \textbf{b} at 250\,eV, and \textbf{c} at 150\,eV. Insets show the normalized D atom REMPI signal as a function of delay time. The flight distance $r$ was calibrated per experimental session using the electron-stimulated desorption falling edge method (see text).
  }
  \label{fig:energy_dep}
\end{figure}

Figure \ref{fig:energy_dep} shows the D velocity distributions obtained from bare gibbsite dried thin film using three different incident electron energies. The inset in the upper right hand corner in each frame shows the time-of-flight data and MB fits whereas the main figures show the velocity distribution obtained by the Jacobian transform, \begin{equation}
    f(v) = S(t) \cdot \frac{(t - t_0)^2}{r}
    \label{eq:jacobian}
\end{equation}
All three irradiation energies used a 3MB part fit with the distribution of slow, medium, and fast ions being 20\%, 47\% and 33\%, as shown in Table \ref{tab:mb_results}. The inset shows data normalized to the largest area compared to time from the end of electron impingement. The raw data is plotted as circles in the inset in black and the fit is presented in gray in both the inset and the main figure. The dashed lines represent the component fits. The interval between the falling edge of the electron gun pulse and the rising edge of the extractor is less than \SI{10}{\nano\second}, making the minimum accessible delay time approximately equal to the laser pulse width itself. D atoms with sufficient velocity to traverse the sample-to-ionization-volume distance $r$ within this window are therefore not detected, resulting in truncation of the high-velocity tail of the distribution. This can be compensated by decreasing $r$ but this limits the signal intensity. This truncation of fast species is shown in the inset which plots the normalized signal intensity v. time. The distributions have a slow portion first observed around 1000 $m s^{-1}$ and a second rise at 3600 $m s^{-1}$. This bifurcation and mixture of slow, medium, and fast neutrals implies two different mechanisms may be present in the release of deuterium atoms. The shape and smoothness of the sample also contributes to a spread in the velocity distribution. The overlapping platelets of gibbsite contribute to the variations of the release angle for the neutral atoms, especially from the exposed edge sites present on gibbsite. With varied components of \({i,j,k}\) in vector space, the time from surface to laser ionization is varied even if two neutrals share the same kinetic energy. This is further effected by the waist of the laser creating varied positions of ionization. 
  
\subsection{Electron Irradiation of Gibbsite: Temperature Dependence}
\begin{figure}[htbp]
  \includegraphics[width=\columnwidth]{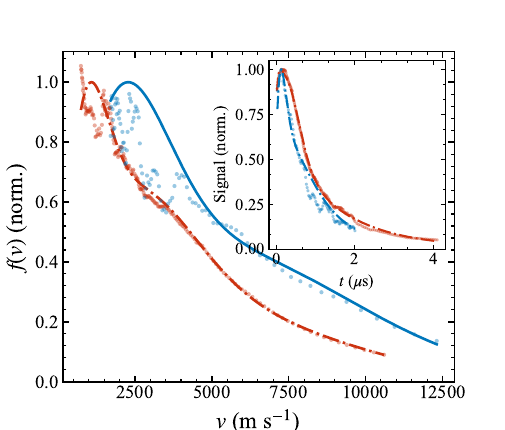}
    \caption{Temperature dependence of the D atom velocity distribution from Gibbsite at 250\,eV irradiation energy. The 15\,K distribution is fitted with a two-component model; signal levels at 15\,K were significantly lower than at 300\,K (see text and Table~\ref{tab:mb_results}). Red and blue curves correspond to 300\,K and 15\,K irradiation temperatures, respectively. The solid blue line and dashed red line show the total Maxwell-Boltzmann fit while the data is represented by data points. Insets show the normalized D atom REMPI signal as a function of delay time.}

    \label{fig:gibb_temp}
\end{figure}

 In order to directly probe the role of substrate temperature, gibbsite was irradiated at 15\,K with 250\,eV electrons. The normalized time-of flight distributions are shown in the inset of Figure \ref{fig:gibb_temp} whereas the main figure depicts the velocity distribution obtained via the previously mentioned Jacobian transform. The overall D neutral yield at 15\,K (blue curve) was considerably reduced (at least 50\%) relative to the yield at 300\,K (red curve) and as shown in Table \ref{tab:signal_summary}. Unlike the data obtained at 300\,K, the low temperature data is well fit using a 2MB distributions with a noticeable lack of the slowest velocity component. As compared to irradiation at 300\,K, the ESD signal remained consistent at 15\,K. Since the D atom yield was significantly reduced and the ESD ions signal was not, D atom formation and desorption was likely intrinsically linked to the electronic properties of the substrate and not the electron flux and terminal site coverage. The medium $968\,K$ and high $7231\,K$ components for cryogenic gibbsite derived from the 2MB model are summarized in Table~\ref{tab:mb_results}. The prevalence of the high velocity component implies that the signal is dominated by nonthermal direct excitations and is likely not due to the presence of adsorbed or trapped water or the presence of ice. Since we are observing the deuterium signal not hydrogen the presence of deposition of ambient water from the chamber with cooling would not contribute to observed REMPI peak.  

\subsection{Electron Irradiation of Gibbsite: The Role of Ice}
To address the possible influence of adsorbed water and ice, electron irradiation experiments were conducted on gibbsite films with nanoscale ice deposited at 15\,K. A low-density amorphous solid water (ASW) coverage of at least 45 monolayers was used to preclude any interaction with the subsurface gibbsite, and deposition at 15\,K prevents the formation of crystalline ice.

\begin{figure}[htbp]
  \includegraphics[width=\columnwidth]{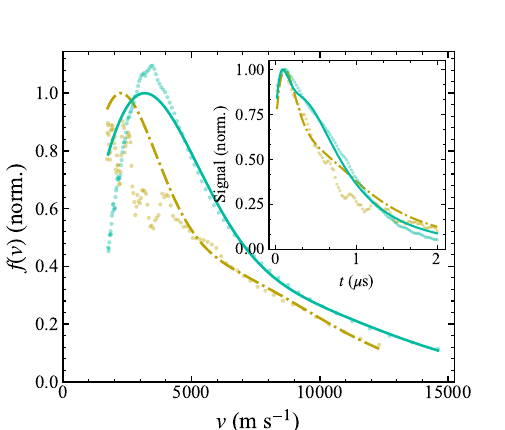}
  \caption{Substrate dependence of the D atom velocity distribution at 250\,eV irradiation energy and 15\,K sample temperature. Gold and Teal data points correspond to bare Gibbsite and \ce{D2O} ice deposited on Gibbsite, respectively. The solid gold line and the teal dash-dot line show the total two-component Maxwell-Boltzmann fits. The inset shows the normalized D atom REMPI signal as a function of delay time. Both conditions share the same flight distance $r = 0.394$\,cm. The \ce{D2O} ice signal is significantly larger than the bare Gibbsite signal (peak area $\sim37-88$ vs. $\sim2$\,arb.\ units), reflecting the higher surface density of deuterium in the ice film.}
  
  \label{fig:substrate}
\end{figure}

The D atom normalized velocity distributions and time-of-flight data are shown in Figure~\ref{fig:substrate}. The velocity distributions were similar in shape as compared to each with \ce{D2O} ASW producing faster D neutral species. A 20-fold increase in integrated signal intensity was observed relative to cryogenic bare gibbsite films. The D neutral signal from \ce{D2O} is faster than that from both room-temperature and cryogenic gibbsite, with the peak occurring at 3178\,\si{\meter\per\second} (107\,meV). %

Under cooling conditions, some deposition of water from the chamber onto the surface is unavoidable. Assuming Langmuir deposition at a backing pressure of \SI{1e-9}{Torr} over a 60\,minute cooling period, less than one monolayer of ambient water would accumulate.

\section{Discussion}
 It is well known that the inelastic scattering of electrons can lead to the stimulated desorption and dissociation of surfaces and adsorbates. This is formally referred to as desorption induced by electronic transitions (DIET). \cite{yatesSynthesisCharacterizationTribophysical} As previously mentioned, when electrons are the impinging particles, the process is typically referred to as ESD. If the incoming energy is above the ionization levels of the target, the primary energy loss channels are ionization followed by multi-electron cascade and relaxation processes. Knotek and Feibelman \cite{knotekIonDesorptionCoreHole1978} developed a model for ESD of covalent and ionic materials, such as metal oxides, that involves interatomic Auger decay following an initial core ionization channel that leads to a reversal of the Madelung potential and ejection of ions. The efficacy of this pathway depends critically on the correlated electron dynamics and eventual hole localization (i.e. self-trapping) probabilities. Full valence metal oxides such as \ce{TiO2}, \ce{ZrO2}, \ce{Fe2O3}, etc. typically have damage cross sections around 10\textsuperscript{-18}\,cm\textsuperscript{2} or lower with \ce{O+} desorption as a major product. Since all metal oxides typically react with water and have M-OH termination sites, \ce{H+},\ce{H3O+}, and sometimes \ce{OH+} are observed. \cite{rothRadiolysisWaterZrO22012, bussiereDynamicsD2Released2006,orlandoElectronStimulatedDesorptionH22003} Though the lattice atoms are not removed efficiently as ions, ionic and neutral fragment may form from adsorbates or termination sites. These desorption pathways typically involve intramolecular Auger decay and two-hole localization within a bond that then undergoes a Couloumb explosion.
In the present experiments, the ESD yield of \ce{O+} and \ce{Al+} is very low indicating that gibbsite is intrinsically very stable to ionization driven radiation damage. This is consistent with previous work with gamma rays \cite{chenEffectsIonizingRadiation2024,huestisRadiolyticStabilityGibbsite2018} and work on a boehmite (\ce{Al-O(OH)}).\cite{jonesElectronStimulatedFormationRelease2022,jonesElectronThermalStimulatedSynthesis2019,jonesEfficientIntermolecularEnergy2020}

 Since samples were drop cast from a gibbsite \ce{D2O} slurry, these surfaces are likely fully terminated by \ce{M-OD} groups which gives rise to the strong \ce{D+} signal. The lack of any other substantial yield of ions other than \ce{D+} indicates that the primary ESD channels do indeed involve two hole states that localize on the terminal hydroxyl groups either on the basal plane or at edge sites. 
\begin{figure*}[htbp]
    \centering
    \includegraphics[width=\linewidth]{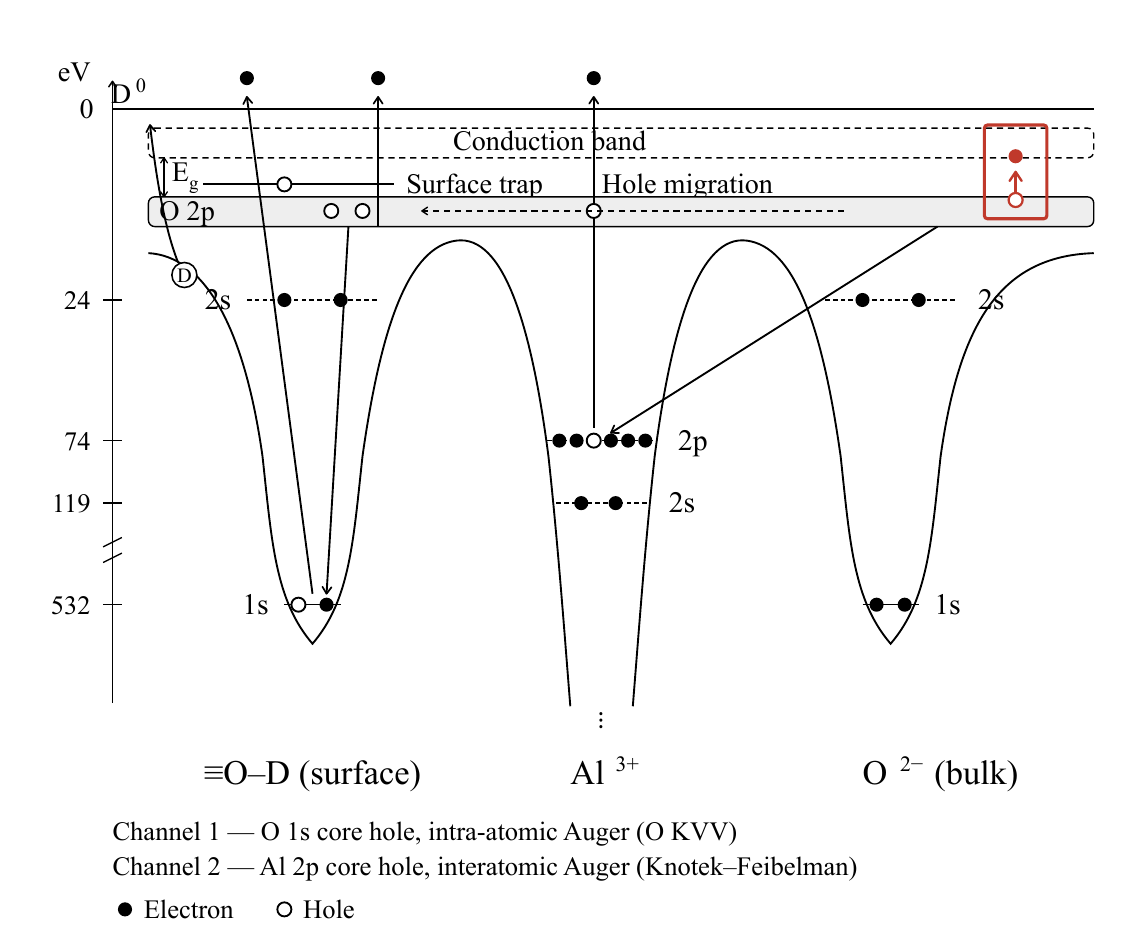}
    \caption{Schematic of the electronic excitation and decay pathways leading to \ce{D} atom desorption from gibbsite. Potential wells for two hydroxyl oxygen sites and a bridging \ce{Al^3+} are shown against a binding energy axis; the axis is broken and compressed between the \ce{Al} 2s and \ce{O} 1s levels. Two primary ionization channels are indicated. In Channel 1, an \ce{O} 1s core hole (532\,eV) decays by intra-atomic Auger emission (\ce{O} KVV), promoting a 2p electron into the core vacancy and ejecting a second 2p electron. In Channel 2, an \ce{Al} 2p core hole (74\,eV) cannot decay intra-atomically, since \ce{Al^3+} carries no valence electrons of its own; the vacancy is instead filled by an electron from a neighboring oxygen 2p orbital, ejecting a second oxygen electron in an interatomic Knotek-Feibelman process. Both channels terminate in holes at the top of the \ce{O} 2p valence band, which migrate and self-trap at a terminal \ce{OD} site slightly above the valence band maximum, where dissociation of the \ce{O-D} bond releases a neutral \ce{D} atom. Filled circles denote electrons and open circles denote holes. The red box indicates the direct creation of holes in the \ce{O} 2p level via electron impingement.}
    \label{fig:auger_mechanism}
\end{figure*}
However, as shown in Figure \ref{fig:auger_mechanism}, single hole states can also be formed and self-trap at the surface leading to bond breakage and, in the case of terminal \ce{OD} site, \ce{D} atom desorption. In view of the limited change in the velocity distributions with varied electron impingement energy shown in Figure \ref{fig:energy_dep}, and the previous observations of a trapped hole in the 2p$^{2}$ oxygen orbital, \cite{kaddissy_radiolytic_2017, kurucParamagneticCentersXrayirradiation1991} it is reasonable to invoke single hole states as the final states leading to D atom desorption. This hole can either be created directly on the terminal site via excitation of the oxygen valence gap. However, in view of the incoming electron energy and penetration depths, it is more likely produced within the sample via a semi-delocalized oxygen vacancy that is stabilized by the lattice structure then followed by migration and self-trapping at the surface terminal hydroxyl group which undergoes dissociation of the terminal D atom. This hole migration is similar to what occurs in other metal oxides.\cite{deskinsIntrinsicHoleMigration2009, tanakaPhotostimulatedIonDesorption2004} 

 A secondary mechanism involving excitation of hole localization on the bridging hydroxyl groups between the lattice plains is also probable. As mentioned, previous reports have observed the creation of stabilized hydrogen atoms after gamma irradiation of boehmite and gibbsite samples clearly demonstrating bulk radiolysis.\cite{huestis_identification_2020} In the subsurface bulk radiolysis case, the rate limiting step for the observation of neutral D atoms within the gibbsite structure is the migration of the D species not the \ce{O-} hole redistribution.\cite{shenDiffusionMechanismsRadiolytic2018} In the case of the present experiments, this would be manifested as a thermalized or quasi-thermalized velocity component. 

The temperature dependence data in Fig. \ref{fig:gibb_temp} addresses the potential role of hole migration and sub-surface dissociation events. Specifically, the decreased D atom signal intensity for the cryogenic systems can be associated with the decreased internal energy of the lattice structure at cryogenic temperatures. This will reduce or prohibit the D species mobility and hole migration and thus decrease the subsequent energy localization and dissociation event. Adsorption of D atoms on the surface is possible at this temperature so this could lead to a reduction of signal primarily from the subsurface products that undergo collisions and have slow velocities. However, this is unlikely to reduce the total yield by measured factor of 20. 

At the low temperature, trapped and physisorbed water can certainly exist. The adsorption and trapping of water on defective metal oxides is known and is typically correlated with a \ce{H3O+} peak in ESD measurements \cite{jonesElectronThermalStimulatedSynthesis2019,orlandoElectronStimulatedDesorptionH22003} and a shift in desorption temperature to about 300\,K for water coupled to the exposed unpassivated positively charged metal ions. \cite{petrikElectronstimulatedReactionsNanoscale2018} 
The lack of any \ce{H3O+} signal during all ESD measurements implies that very little residual or adsorbed water remained in or on the samples. 

Since we could not completely rule out the presence of trapped or strongly chemisorbed water, as discussed above, we carried out experiments that involved intentional water dosing and multilayer ice growth. Though not shown in the normalized data, the D yield from ice shown in Fig.~\ref{fig:substrate} and in Table~\ref{tab:signal_summary} is significantly higher than the yield from bare gibbsite at both 300\,K and 15\,K. More importantly, the velocity distribution is broader for the ice with a peak velocity at 3178\,\si{\meter\per\second}. This is in close agreement with previous work on the ESD of D neutrals from \ce{D2O} ice where they found a peak velocity of 85 \,meV which corresponds to 2853\,\si{\meter\per\second}. \cite{kimmelLowEnergy5120EV1995} The difference in peak location between the value calculated in this work and previous work can be attributed to the varied experimental geometry. In the case of work by Kimmel and Orlando \cite{kimmelLowEnergy5120EV1995}, the D atom yield resulted from the decay of the lowest exciton level of the water which formally involves occupancy of the dissociative \(3s4a_1\) level of water.

\subsection{Geometric and electronic structure}

\begin{figure*}[htbp]
  \includegraphics[]{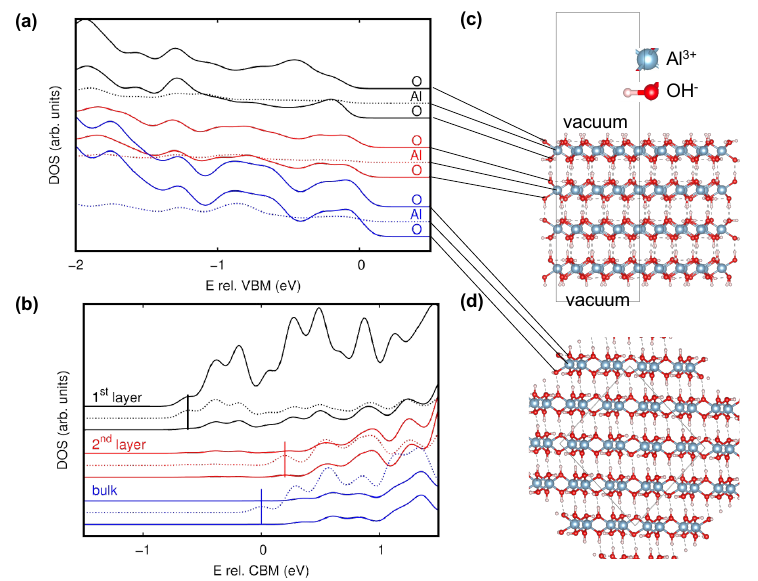}
  \caption{
    Sums of densities of electronic states projected onto \ce{O} 2p and \ce{Al} 3s orbitals near the valence (a) and conduction (b) band edges in a 4-layer (001)-oriented slab model (c) and bulk gibbsite (d). The sums extend over layers of O (solid lines) and \ce{Al} (dotted lines) in the surface octahedral layers (black), the subsurface octahedral layers (red), or the bulk (blue), in which all layers are equivalent. The highest occupied states near the valence band edge have similar positions in energy and weight from the \ce{O} 2p orbitals in the near-surface and bulk models, while the lowest unoccupied orbitals (marked by vertical lines in panel (b)) are significantly lower in energy in the vacuum-facing surface layer.
  }
  \label{fig:gibbsite_dos}
\end{figure*}

Gibbsite is a fully hydroxylated aluminum hydroxide whose structural hydroxyl groups govern the interlayer interactions and stacking patterns. Figure \ref{fig:gibbsite_dos} shows that the layers of octahedral co-ordinated \ce{Al} cations are between layers of pseudo-hexagonally close-packed hydroxyl groups. These layers stack in the z-direction with oxygen atoms in each layer directly opposing those in adjacent sites. \cite{frostVibrationalSpectroscopyDehydroxylation1999} This leads to six structurally distinct \ce{OH} groups within each layer, three in the interlayer direction and three in the intra-layer direction. Though these interlayer \ce{OH} groups did undergo dissociation during \ce{^{60}Co} irradiation, Wang et. al.\cite{wangRoleSurfaceHydroxyls2020} used interface-selective vibrational sum frequency generation to show that damage involving the surface hydroxyl groups was significantly higher than the “internal” structural \ce{-OH}. These authors further noted a lack of trapped H but found trapped electrons, \ce{O.}, \ce{O2-}, possibly F-Centers, and disruption of the surface structure. 

\begin{figure}[htbp]
  \includegraphics[width=\columnwidth]{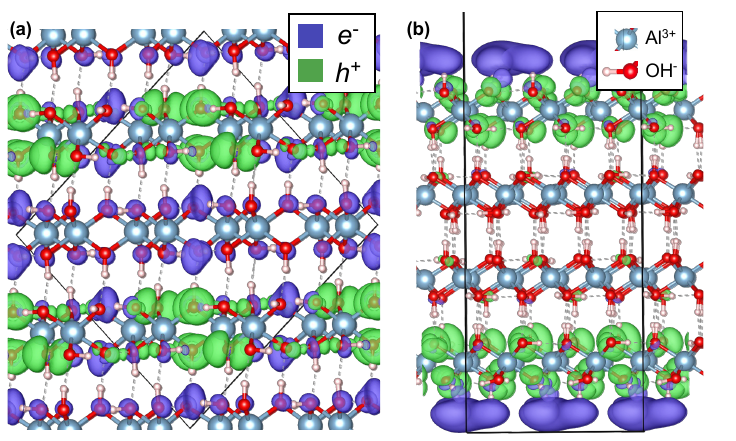}
  \caption{
    Density isosurfaces at 0.005 electrons/\AA{}$^3$ of the empty lower (green, hole) and occupied upper (blue, electron) orbitals from the half-occupied orbitals in the spin-constrained polaron models of bulk (a) and (001) terminated 4-layer slab (b) gibbsite. Thin black lines mark the periodic cells of the simulations, which are identical to those depicted in Figure \ref{fig:gibbsite_dos}. In the slab model, the electron orbital is mostly carried by the terminal hydroxyl groups on the upper and lower vacuum-facing surfaces, and the hole primarily resides in the O sites in surface octahedral layers. In the bulk model, which does not have any surfaces, both carriers are extended throughout the cell. 
  }
  \label{fig:exciton_orbitals}
\end{figure}

As shown in Fig.~\ref{fig:gibbsite_dos}, the structural and surface hydroxyls have different configurations which can affect the disorder, electronic structure and carrier (hole and electron) transport properties. In an effort to address the radiolytic yield of molecular hydrogen from gibbsite vs boehmite, first-principles electronic structure calculations employing hybrid density functionals have been done for bulk gibbsite. \cite{prange_electronic_2018} These calculations reported a band-gap energy and an energy loss function that were compared to X-ray photoelectron and electron energy loss measurements. The results demonstrated that low energy electrons are isotropically mobile, while holes in the valence band are likely constrained to move laterally within structural layers rather than perpendicular to the structural layers.
To investigate the role of the surface in excitation dynamics, we have conducted additional electronic structure calculations using slab models exposing the dominant (001) facet exhibited by gibbsite nanoparticles and reference calculations of the gibbsite bulk. The band gap of the extended bulk using these approximations (PBE-D3) is 5.43\,eV, significantly underestimating the true value ($\approx$ 8\,eV\cite{prange_electronic_2018}) as is expected for semi-local exchange-correlation functionals. However, we believe relative differences between states of the same occupation are faithfully captured in these calculations. The computed electronic structure of the ground state shows that the surface has only a minor effect on the shape and position of the occupied states, but the lowest-lying empty states near the surface are stabilized by about 0.6\,eV relative to the extended bulk (Fig. \ref{fig:gibbsite_dos}b). To simulate the interaction of excited carriers, spin triplet states were calculated for the relaxed ground state geometry, and then the atomic coordinates were relaxed while constrained to the triplet spin state. The difference in energy between the singlet and triplet states estimates the excitation energy to create a triplet exciton. The energy gained by structural relaxation on the triplet potential energy surface gives an estimate of the exciton trapping energy. Within the limitations of semilocal DFT, the resulting orbitals serve as qualitative models for free and trapped excitons, and their relative energetics near the surface and in the bulk models give insight into the energetic landscape of excited carriers in nanosized gibbsite. In bulk gibbsite, we obtain an excitation energy of 5.77\,eV in the ground state geometry which relaxes 0.04\,eV in the triplet excited state. For the (001) slab model, the excitation energy is 5.12\,eV in the ground state, and the relaxation energy is 0.17 eV. The resulting orbitals are shown in Figure \ref{fig:exciton_orbitals}. As noted above, PBE-D3 is expected to overestimate the energy (i.e. underestimate the stability) of localized states, so the trapping energies are likely underestimated compared to the true values, and the orbitals shown in Figure \ref{fig:exciton_orbitals} are likely more extended. The general picture that emerges at this level of theory is that a single electron-hole pair exists as a weakly bound large polaron\cite{itoh2001materials} (exciton) that sees a potential well near surfaces that encourages coalescence in the surface layer and trapping there.
Some indirect experimental evidence for this picture can be inferred from the work of Jiang and Spence, {\cite{JIANG2011860}} who observed an electron energy loss feature near 9\,eV that was suppressed over time by exposure to the 200\,keV  electron probe beam of intensity 59\,pA/cm$^2$. These observations are consistent with assignment of the 9\,eV feature to direct excitation of final states similar to Figure~\ref{fig:exciton_orbitals}b by the probe beam that are disrupted by knock-on removal or secondary electron stimulated removal of the terminal H over time.

\section{Conclusion}
The electron-stimulated desorption of neutral D atoms from gibbsite (\ce{Al(OD)3}) nanoplatelets and amorphous \ce{D2O} ice was investigated using $2+1$ REMPI time-of-flight mass spectrometry at 15 and 300\,K. The neutral D atom channel was examined and its translational energy distribution was characterized as a function of both incident electron energy and substrate temperature.
At 300\,K, irradiation at 540, 250, and 150\,eV produced similar D atom velocity distributions, each well described by a three-component Maxwell-Boltzmann model. The insensitivity of the distributions to incident energy across the O\,1$s$ core-level threshold indicates that the desorption dynamics are governed by the final-state relaxation of localized holes rather than by the initial excitation channel. Cooling to 15\,K reduced the D atom yield by at least 50\% and eliminated the lowest-temperature (slowest) component, while the ESD ion signal was unchanged. This decoupling of the neutral yield from the ion signal ties D atom formation to the electronic and transport properties of the substrate rather than to electron flux or terminal-site coverage.
We identify two channels contributing to neutral D atom production. In the first, a hole created within the sample migrates to a terminal hydroxyl group, where it self-traps and drives dissociation of the O--D bond, releasing a D atom from the surface. In the second, subsurface radiolysis forms D atoms within the structure that must then diffuse to the surface to desorb, giving rise to the lower-velocity component of the distribution. Both channels depend on thermally activated hole migration and proton mobility, and both are correspondingly suppressed at cryogenic temperatures, consistent with the observed loss of yield and loss of the slowest velocity component at 15\,K. Spin-constrained DFT calculations support this picture, showing that exciton states and excited holes are stabilized near the (001) surface and that energy localizes preferentially at terminal hydroxyl sites.
Deuterated amorphous solid water films produced roughly 20 times the D atom signal of bare gibbsite and a distinctly hotter, broader velocity distribution, reflecting the higher deuterium surface density and the different bonding environment of bulk ice relative to the terminal hydroxyls of gibbsite. The D atom yields and distributions from gibbsite are therefore not attributable to adsorbed or trapped water. Taken together, these results establish hole transport to terminal hydroxyl sites as the rate-limiting step for nonthermal D atom release from aluminum hydroxide surfaces. Quantifying the associated \ce{D2} yields, which are not detected in the present measurements, remains the subject of future work.

\begin{acknowledgments}
This research was supported by IDREAM (Ion Dynamics in Radioactive Environments and Materials), an Energy Frontier Research Center funded by the U.S. Department of Energy (DOE), Office of Science, Basic Energy Science (BES) under FWP 68932. The electron irradiation work was carried out at Georgia Tech. The materials synthesis portion of this research was performed using the Environmental Molecular Sciences Laboratory, a national scientific user facility sponsored by the DOE Office of Biological and Environmental Research and located at Pacific Northwest National Lab (PNNL). PNNL is a multiprogram national laboratory operated for DOE by Battelle Memorial Institute under Contract DE-AC0576RL0-1830.
\end{acknowledgments}
See Supplementary Information for additional data.
\bibliography{references}

\end{document}